\documentclass[aps,twocolumn, pra,superscriptaddress, nofootinbib,longbibliography]{revtex4-2}

\usepackage[utf8]{inputenc}
\usepackage{amsmath, amssymb}
\usepackage{graphicx}
\usepackage{mathrsfs} 
\usepackage[colorlinks=true, allcolors=blue]{hyperref}
\usepackage{braket} 
\usepackage{graphicx}
\usepackage{subcaption}
\usepackage{xcolor}
\usepackage[justification=raggedright,singlelinecheck=false]{caption}

\begin{document}

\title{Rydberg atom parity gate based on dark state resonances}

\author{Sinchan Snigdha Rej}
\email{sinchansrej@gmail.com}

\author{Snigdhadev Ray}
\email{snigdhadevray@gmail.com}

\author{Bimalendu Deb}
\email{bdiacs@gmail.com}

\affiliation{School of Physical Sciences, Indian Association for the Cultivation of Science, Jadavpur, Kolkata 700032, India}


\begin{abstract}
 Quantum computation (QC) and digital quantum simulation (DQS) essentially require two- or multi-qubit controlled-NOT or -phase gates. We propose an alternative pathway for QC and DQS using a three-qubit parity gate in a Rydberg atom array. The basic principle of the Rydberg atom parity gate (RPG) is that the operation on the target qubit is controlled by the parity of the control qubits. First, we discuss how to construct an RPG based on a dark state resonance. We optimize the gate parameters by numerically analyzing the time evolution of the computational basis states to maximize the gate fidelity. We also show that our proposed RPG is extremely robust against the Rydberg blockade error. To demonstrate the efficiency of the proposed RPG over the conventional CNOT or CZ gate in QC and DQS, we implement the Deutsch–Jozsa algorithm and simulate the Ising Hamiltonian. The results show that RPG can be a better substitute of the CNOT gate to yield better results, as it decreases the circuit noise by reducing circuit depth.

\end{abstract}

\maketitle
\section{introduction} In recent years, neutral atom quantum computation has become one of the most important topics of research in physics for its inherent power of scalability. Two- and multi-qubit controlled-NOT (CNOT) and -phase (CZ) gates are the basic building blocks for simulating quantum algorithms, quantum error-correcting codes, and quantum communication protocols. Scientists have proposed and demonstrated several techniques to realize quantum gates based on Rydberg blockade \cite{PhysRevLett.104.010503, PhysRevA.92.022336, Shi2017RydbergGatesFree,PhysRevA.104.012615,PhysRevA.110.062618} and antiblockade \cite{PhysRevA.95.022319, Wu2021AntiblockadeSWAP,Wu2021ResilientRydberg,  Li2023HighToleranceAntiblockade}, techniques using dark states \cite{PhysRevLett.102.170502, Petrosyan2017, PhysRevLett.129.200501, RejDeb2025_ToffoliCnNOT, Rej2025_NeutralAtomGeometric}, time-optimal global pulse \cite{PhysRevLett.123.170503,JanduraPupillo2022TimeOptimalRydberg,Evered2023HFPE}, analytically shaped pulse \cite{PhysRevA.94.032306}, adiabatic and non-adiabatic passage \cite{Rao2014_RobustRydbergAdiabatic, Beterov2016_StarkTunedForster, Saffman2020_SymmetricRydbergCZ, Wu2018_RydbergAdiabaticPhaseControl, Liao2019_GeometricRydbergSTA, Zhou2020_STA_CZ_Rydberg, Rej2025_NeutralAtomGeometric, PhysRevResearch.2.043130}, etc. Many of the applications of QC rely on measuring the parity of the control qubits by applying multiple two-qubit CNOT or CZ gates. But multiple layers of two-qubit gates increase the circuit depth, which might degrade the outcome. To get rid of the noise due to circuit depth, a new type of multi-qubit quantum gate called the parity-controlled gate, or parity gate has been proposed \cite{Su2020_RPM,Zheng2020_RPM_optIE}. The basic principle of this gate is that the operation on the target qubit depends on the even or odd parity of the control qubits, unlike the conventional CNOT or CZ gates. In the recent years, various techniques to realize parity gates and their potential applications are proposed, for example, four-qubit parity gate implementation \cite{Dlaska2022}, the quantum Fourier transform and quantum
addition by parity encoding \cite{Fellner2022}, quantum teleportation using nondestructive Rydberg parity meter \cite{Su2020_RPM}, implementation of RPG by optimized inverse engineering \cite{Zheng2020_RPM_optIE}, quantum error detection by parity-controlled gate \cite{Guo2024_parity_control}, multi-qubit parity gates in various configuration \cite{Kazemi2025_multi_qubit_parity}, etc.

In this paper, we propose a protocol to realize a three-qubit RPG using dark state resonances. Let us consider a linear configuration of three atomic qubits where a target qubit is placed at the mid-point between two control qubits as schematically shown in Fig.\ref{fig1}. The protocol begins with the application of simultaneous $\pi$ pulses on the control atoms. Subsequently, the target atom undergoes two-photon Raman transitions, which are controlled by the parity of the two control qubits. When the control atoms are in the even parity, the dark state condition holds good preventing any population transfer in the target atom. On the other hand when the control atoms have odd parity, the dark state condition breaks down leading to population inversion in the target atom. This is followed by the application of another set of simultaneous $\pi$ pulses on the control atoms to complete the protocol.

The gate parameters are optimized via numerical analysis of the time evolution of computational basis states to maximize fidelity. We achieve a gate fidelity of $99.35\%$ for gate operation time 0.27 $\mu$s in the presence of spontaneous decay of the Rydberg states by solving the master equation using QuTip \cite{qutip2012}. We further show that our proposed gate is robust against the Rydberg blockade error. To establish the supremacy of our proposed RPG over the widely used CNOT or CZ gates, we implement the Deutsch–Jozsa algorithm and digitally simulate the Ising Hamiltonian using Qiskit \cite{qiskit}.We demonstrate that a single RPG can replace multiple layers of two-qubit gates, leading to improved outcomes in QC and DQS through the reduction of circuit depth, thereby minimizing circuit noise.

The paper is organised in the following way. In the section \ref{RPM} we discuss the gate protocol in detail. Then, in the section \ref{RD}, we show our numerical results of the evolution of computational states by varying gate parameters, analysis of gate fidelity with respect to gate time, robustness against the blockade error, and the application of RPG in QC and DQS. Then we conclude in section \ref{cncl}.

\section{Three-qubit RPG}\label{RPM}

 \begin{figure}[t]
     \centering
    \includegraphics [width=8.5cm]{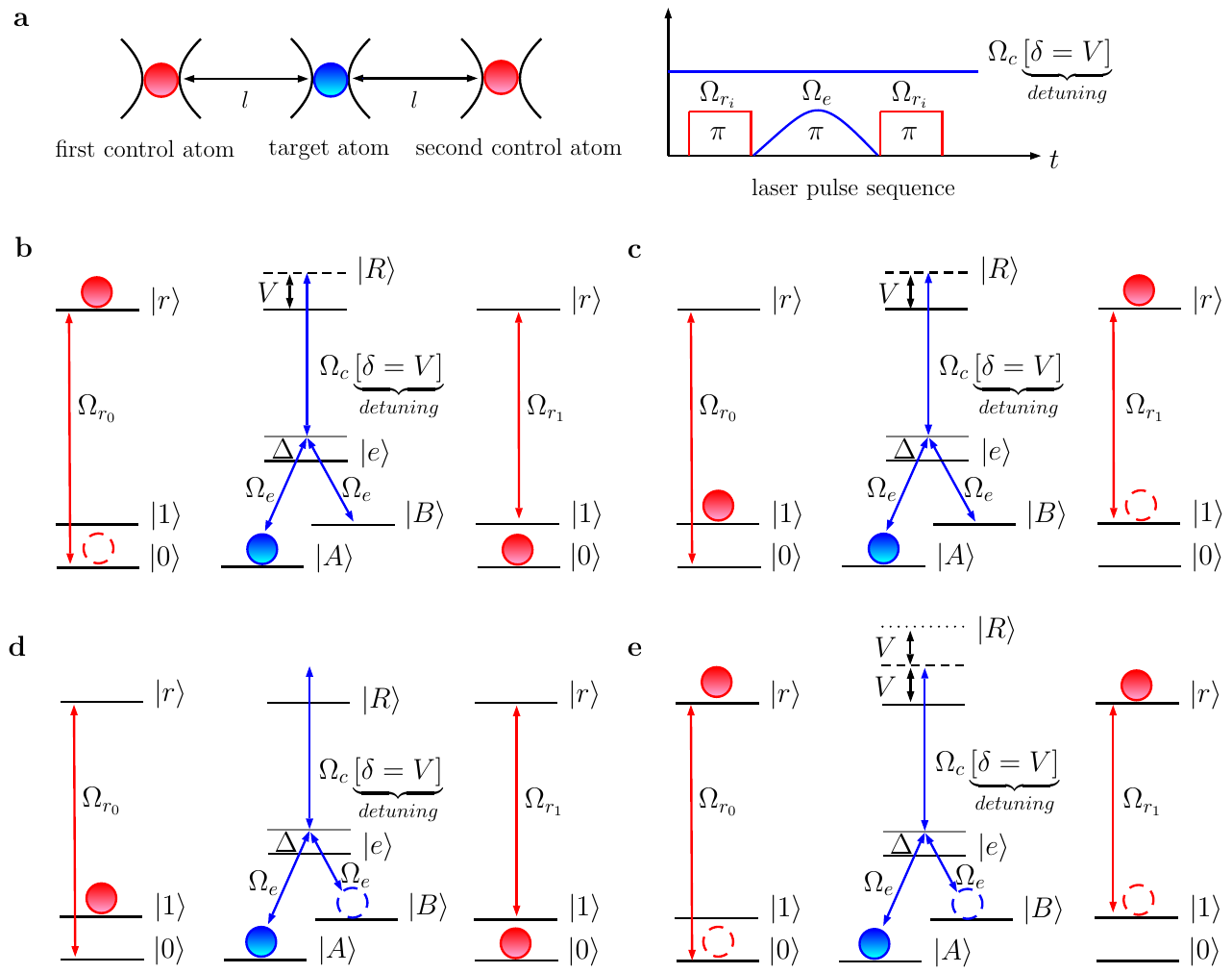}
     \caption{(a) Three atoms are trapped in three different optical traps at a fixed inter-atom spacing $l$, and the pulse sequence to complete the gate protocol is shown. (b) and (c) show that any evolution in the target atom is forbidden when the parity of the control qubits is even. (d) and (e) show that the evolution $\ket A \longleftrightarrow \ket{B}$ happens in the target atom when the control qubits are in odd parity. }
     \label{fig1}
 \end{figure}
Let us consider that the two control atoms and one target atom are confined in three different optical traps with a distance $l$ between the trap centres, as shown in Fig.\ref{fig1}a. Two hyperfine ground states $\ket 0$ and $\ket 1$ are considered as the two qubit states, and $\ket r$ is the Rydberg states of both the control atoms. For the first control atom, $\ket 0$ is coupled to $\ket r$ with a Rabi frequency $\Omega_{r_0}$ while for the second control atom, $\ket 1$ is coupled to $\ket r$ by another laser with Rabi frequency $\Omega_{r_1}$. For simplicity we choose $\Omega_{r_0}=\Omega_{r_1}=\Omega_r$. Similarly, two hyperfine ground states $\ket A$ and $\ket B$ are considered as the qubit states for the target atom. $\ket A$ and $\ket B$ are in two-photon off-resonance with the corresponding Rydberg state $\ket R$ of the target atom via an intermediate state $\ket e$. $\ket A$ and $\ket B$ are off resonantly coupled with $\ket e$ with a pair of soft Raman $\pi$ pulses with Rabi frequency $\Omega_e(t)$ and blue detuing $\Delta \gg \gamma_e$ to minimize spontaneous decay from $\ket e$. $\ket e$ is also off-resonantly coupled with $\ket R$ with a Rabi frequency $\Omega_c$ and blue detuning, $\delta$ as shown in Fig.\ref{fig1}.

\paragraph{Case-I (even parity)}: When both control atoms are at $\ket{0}$, after applying the first $\pi$ pulse, only the first one will be evolved to $\ket r$ while the second one will remain unchanged. It will result in a shift of $V$ in the state $\ket R$ of the target atom due to strong Rydberg-Rydberg interaction. Now, after applying the soft Raman $\pi$ pulse, the target atom will follow the Hamiltonian ($\hbar =1$)
 \begin{equation}  
 	\begin{aligned}
     \mathcal{H}_1(\delta,t) =&  \Omega_e(t)/2(\ket{A}\bra{e}+\ket{B}\bra{e})- \Delta\ket{e}\bra{e}\\ &+ \Omega_c/2 \ket{e}\bra{R}e^{i\delta t} + V\ket{R}\bra{R}+\text{H.c.}.
     	\end{aligned}
 \end{equation}
Applying the unitary transformation $e^{-iV\ket{R}\bra{R}}$ and putting $\delta=V$ we get
\begin{equation} 
	\begin{aligned} 
     \mathcal{H}_1(t) =&  \Omega_e(t)/2(\ket{A}\bra{e}+\ket{B}\bra{e}) - \Delta\ket{e}\bra{e}\\ &+ \Omega_c/2 \ket{e}\bra{R}+\text{H.c.}.
     \label{Ht1}
     \end{aligned}
 \end{equation}
The above Hamiltonian has two dark states $\ket{\mathcal D_1}=1/\sqrt{2}(\ket{A}-\ket{B)}$ and $\ket{\mathcal D_2}=(1+y^2)^{-1/2}[1/\sqrt{2}(\ket{A}+\ket{B}-y\ket{R})]$ where $y=\sqrt{2}\Omega_e(t)/\Omega_c $. Satisfying the condition $\Omega_c/\Omega_e>2 $, the target qubit follows the dark state $\ket {\mathcal D} =\frac{1}{\sqrt2}(\ket {\mathcal{D}_1} +\ket {\mathcal{D}_2)} $ during evolution, resulting in the transformation $\ket{A} \longrightarrow \ket{A}$,
as depicted in Fig.\ref{fig1}b. When both control atoms are in $\ket 1$, similar arguments hold good and no time evolution occurs in the target atom as a result (see Fig.\ref{fig1}c). 

\paragraph{Case-II (odd parity)}: When the first control atom is in $\ket 1$ and the second one is in $\ket 0$, none of them will be evolved to $\ket r$ upon application of the first $\pi$ pulse. Consequently, there is no shift in $\ket R$, and the dark state condition continues to hold good and the dynamics is governed by the Hamiltonian
 \begin{equation}  
 	\begin{aligned}
     \mathcal{H}_2(\delta,t) =&  \Omega_e(t)/2(\ket{A}\bra{e}+\ket{B}\bra{e}) - \Delta\ket{e}\bra{e}\\ &+ \Omega_c/2 \ket{e}\bra{R}e^{i\delta t}+\text{H.c.}.
     \end{aligned}
 \end{equation}
From the previous case we know $\delta=V$ and for $V>\Omega_c^2/(4\Delta)$, the dark state condition breaks down completely to allow the evolution $\ket{A}\longleftrightarrow\ket{B}$. For the initial state, when the first control atom is in $\ket 0$ and the second one is in $\ket 1$, after applying the first $\pi$-pulse, $\ket r$ of the both control atoms will be populated, and as a result, $\ket R$ will experience a shift of $2V$. The target atom will evolve following the Hamiltonian, 
 \begin{equation}  
 	\begin{aligned}
     \mathcal{H}_3(\delta,t) =& \Omega_e(t)/2(\ket{A}\bra{e}+\ket{B}\bra{e})- \Delta\ket{e}\bra{e}\\ &+ \Omega_c/2 \ket{e}\bra{R}e^{i\delta t} +2V\ket{R}\ket{R}+\text{H.c.}.
     \end{aligned}
\end{equation}
Applying the unitary transformation $e^{-i2V\ket{R}\ket{R}}$ and taking $\delta=V$ the Hamiltonian becomes 
 \begin{equation}  
 	\begin{aligned}
     \mathcal{H}_3(V,t) &=  \Omega_e(t)/2(\ket{A}\bra{e}+\ket{B}\bra{e})+ \Omega_c/2 \ket{e}\bra{R}e^{-iV t}\\& +\text{H.c.}.
     \end{aligned}
\end{equation}
Using a similar condition just like the above case, one obtains the evolution $\ket A \longleftrightarrow \Ket B$ by breaking the dark state condition in the target atom. Next we apply the second $\pi$ pulse on the control atoms to complete the gate operation.
 
\section{Results and discussions} \label{RD}
\begin{figure} 
    \centering
    \includegraphics[width=4.2cm]{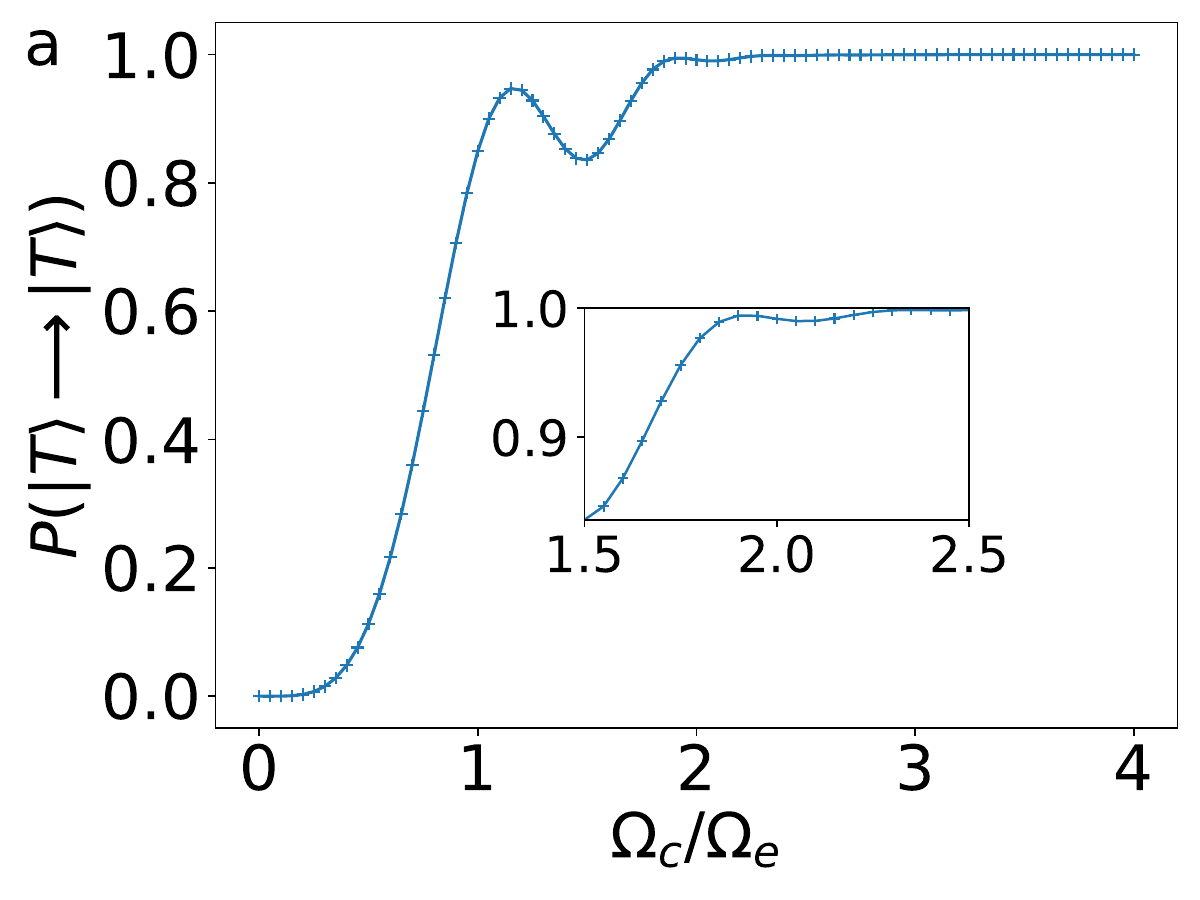}
    \includegraphics[width=4.2cm]{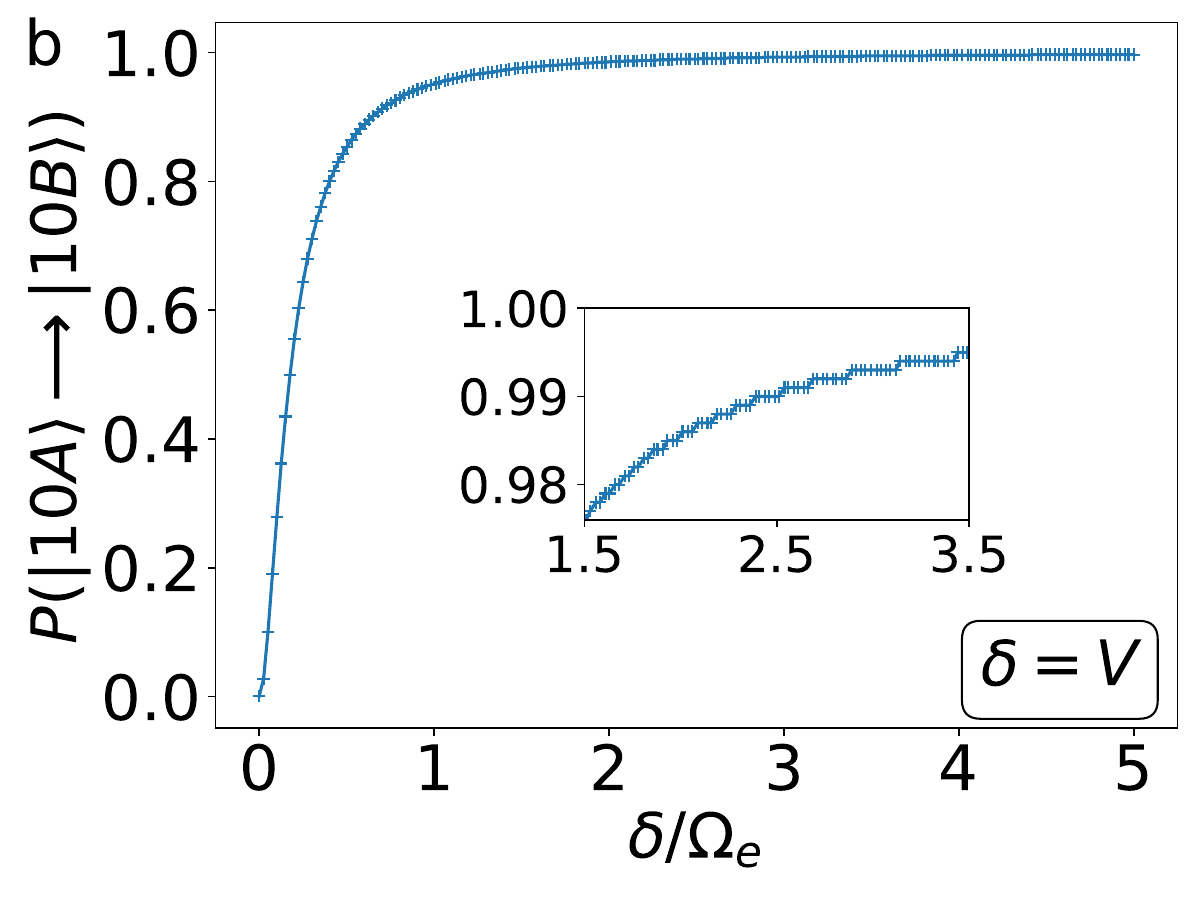} \\
    \includegraphics[width=4.2cm]{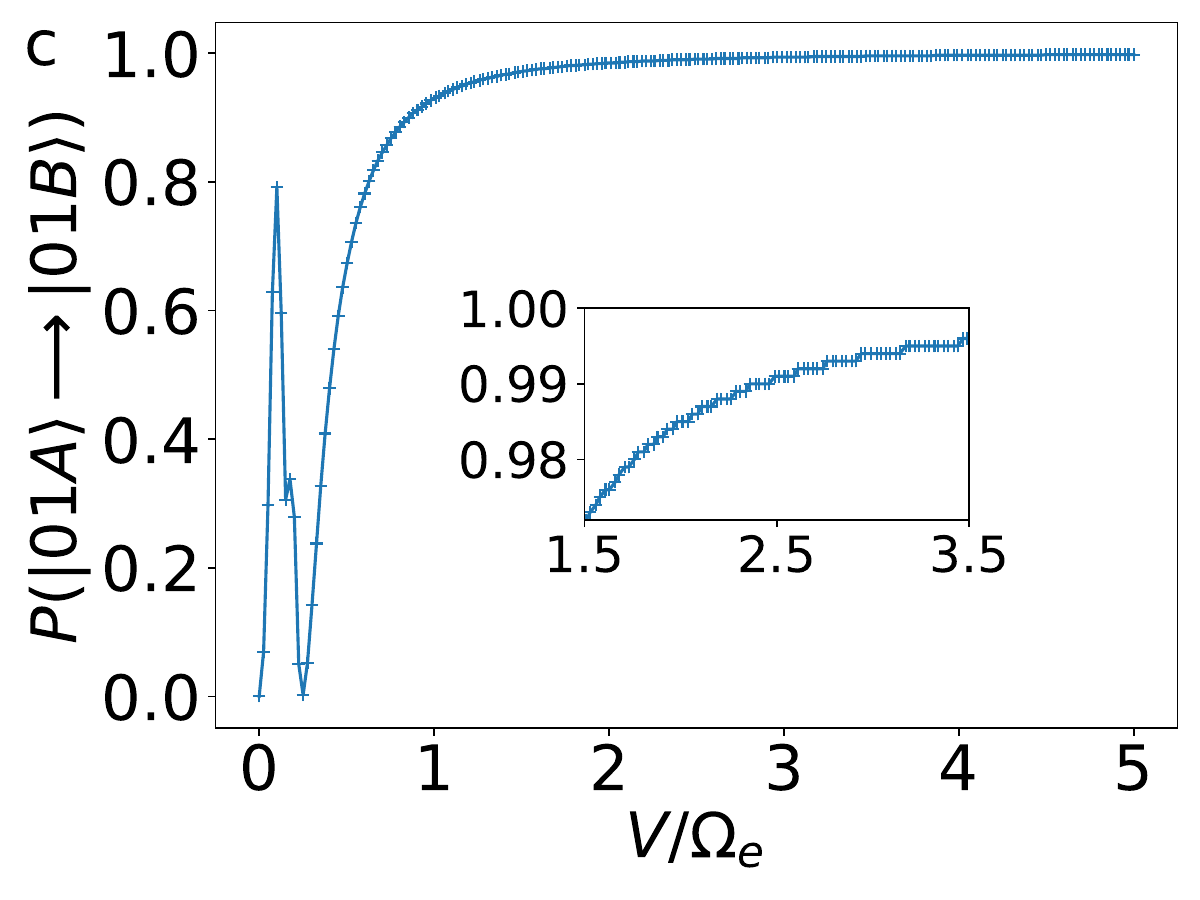}
    \includegraphics[width=4.2cm]{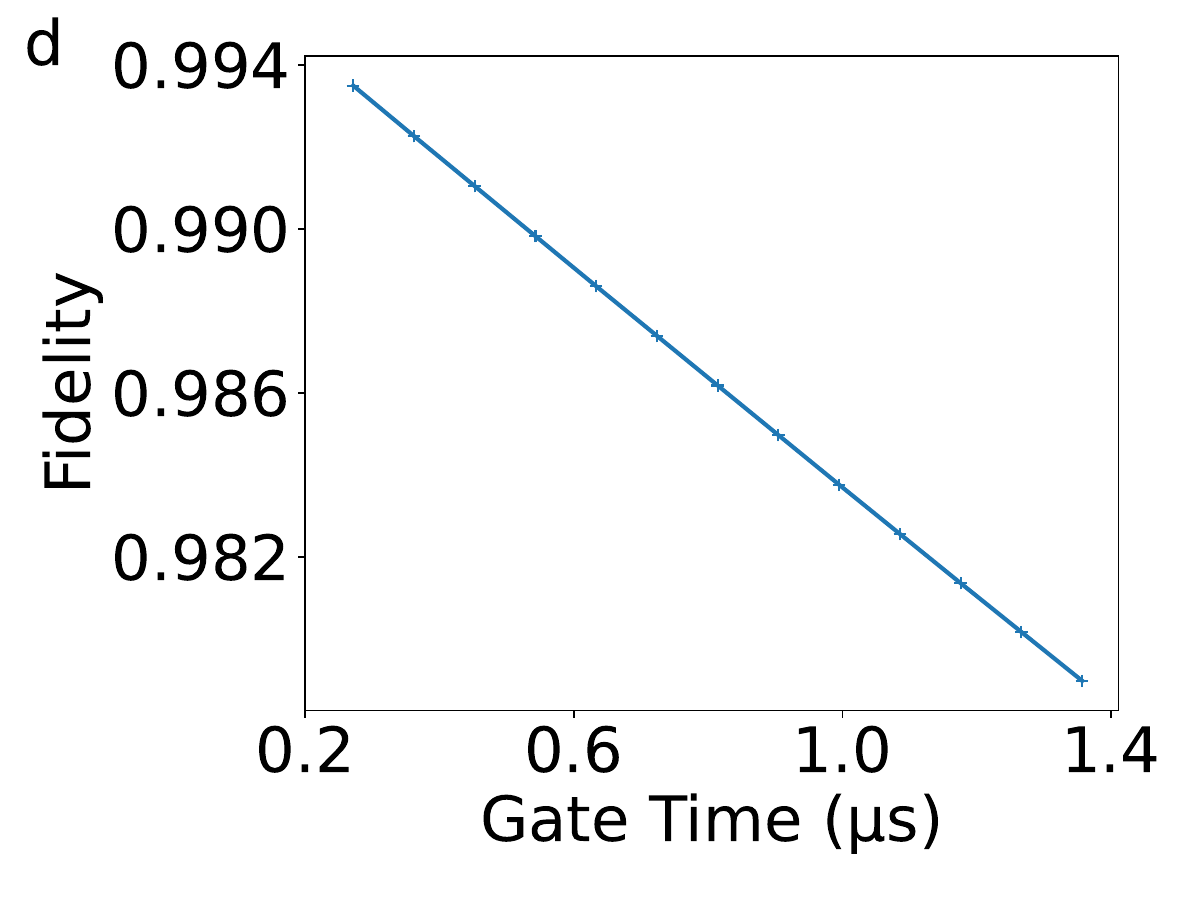}
    \caption{(a) For even parity of the control qubits, population transfer in the target atom does not happen for $\Omega_c/\Omega_e>2$. (b) For odd parity($\ket{10}$), population transfer in the target qubit happens as $\delta$ increases. (c) For odd parity($\ket{01}$), population transfer in the target qubit happens for increasing $V$. (d) The variation of average gate fidelity is plotted as a function of total gate time.   }
    \label{fig2}
\end{figure}
To implement our proposed RPG, we first optimize the system parameters by analyzing the population dynamics. The Rabi frequency of the laser that couples the ground state of the control atoms and $\ket{r}$ is chosen as $\Omega_r = 3\Omega_e$, so the duration of the first $\pi$ pulse is $\mathscr T_1 = \pi/3\Omega_e$. For the Raman transition, we apply a smooth adiabatic pulse, 
\begin{equation}
    \Omega_e(t) = (\Omega_e/2)(1 - \cos(2\pi t/\mathscr T_2))
\end{equation}
with a pulse area of $3\Omega_e^2\mathscr T_2/(16\Delta)$. For a $\pi$ pulse, this yields 
\begin{equation}\label{ET2}
   \mathscr T_2 = 16\pi\Delta/(3\Omega_e^2).
\end{equation}
 To minimize decay of $\ket e$, we choose $\Delta \gg \gamma_e$, where $\gamma_e = 1/\tau_e$. Since the control atom is in $\ket{r}$ and decays at a rate $\gamma_r = 1/\tau_r$, the Raman pulse must be much faster, i.e. $\mathscr{T}_2 \ll \tau_r$. The time duration for the third $\pi$ pulse is 
\begin{equation}\label{ET1T2}
    \mathscr T_3 = \mathscr T_1 = \pi/\Omega_r.
\end{equation}
 The total gate operation time is $\mathscr T_{Tof}=(\mathscr T_1+\mathscr T_2+\mathscr T_3).$

Now, let us examine the population of computational basis states after the entire gate operation is over by varying the computational parameters. Fig.~\ref{fig2}a illustrates the probability of $\ket{T}\longrightarrow\ket{T}$, $\ket T $ being defined as the target qubit state, that is, the probability of no change in the target state, when the parity of control qubits is even. For this plot, $\Omega_c/\Omega_e>2$, this figure shows that the population transfer is blocked with $>99\%$ probability. Fig.~\ref{fig2}b shows that increasing $\delta$, the probability of population transfer from $\ket{10A}$ to $\ket{10B}$ increases, when $\Omega_c/\Omega_e =2.5$ and $\Delta/\Omega_e=10$ are kept fixed . Similarly for large $V$, population transfer from $\ket{01A}$ to $\ket{01B}$ happens, as depicted in Fig. \ref{fig2}c.
\subsection{Fidelity}
 For numerical illustration, we consider the $^{133}$Cs atom. We choose two ground state hyperfine sublevels of $6S_{1/2}(F=3,4)$ as both control and target qubit states. $126S_{1/2}$ state with lifetime 540 $\mu$s \cite{Saffman2020_SymmetricRydbergCZ} is chosen as the Rydberg states for both control and target atoms. We choose the intermediate state ($\ket e$) as $7P_{1/2}$ state with the lifetime of 165 ns \cite{Toh2019_7pCesiumLifetimes}.

The average gate fidelity can be evaluated using a trace-preserving, operator-based expression \cite{nielsen2002simple}:
\begin{equation}\label{Efid}
   \bar{F}(\hat{\mathcal{O}},\epsilon)
   = \frac{\sum_j \mathrm{Tr}\!\left[\hat{\mathcal{O}}\,\hat{\mathcal{O}}_j^\dagger\,\hat{\mathcal{O}}^\dagger\,\epsilon(\hat{\mathcal{O}}_j)\right] + d^2}{d^2(d+1)} ,
\end{equation}
where $\{\hat{\mathcal{O}}_j\}$ represents the complete set of tensor products of $N$-qubit Pauli operators. 
For instance, in the three-qubit case, the operator set is $
(\hat{I}\otimes\hat{I}\otimes\hat{I},\; \hat{I}\otimes\hat{I}\otimes\hat{\sigma}_x,\; \ldots,\; \hat{\sigma}_z\otimes\hat{\sigma}_z\otimes\hat{\sigma}_Z).
$ Here, $\hat{\mathcal{O}}$ denotes the ideal RPG, 
$\epsilon$ is the trace-preserving quantum operation realized by the proposed gate implementation, 
and $d = 2^N$ denotes the Hilbert space dimension of an $N$-qubit system. In the presence of the spontaneous emission of Rydberg atoms, the evolution of the whole system can be governed by the master equation
\begin{equation}
    \dot{\hat{\rho}}=-i[\hat H, \hat \rho]+ \sum_{k=1}^2\sum_{i=0}^1\mathscr{D}_1[\sigma_{i,k}]\hat{\rho}+\mathscr{D}_2[\alpha]\hat{\rho}+\sum_{j=A}^B\mathscr{D}_3[\beta_j]\hat{\rho},
\end{equation}
where $\mathscr {D}[\hat a]\hat \rho=a\hat \rho\hat a^\dagger-(\hat\rho\hat a^\dagger \hat a+\hat{a}^\dagger\hat a\hat \rho)/2$. The decay channels are defined as  $ \sigma_{i,k}=\sqrt{\gamma_r/2}\ket{i}_k\bra{r}$ for the $k$-th control atom, $\alpha=\sqrt{\gamma_R}\ket{e}\bra{R} $ and $\beta_j=\sqrt{\gamma_e/2}\ket{j}\bra{e}$ for the target atom.  We analyze the average gate fidelity by varying the gate operation time in the Fig.\ref{fig2}d and show that for gate operation time $0.27\ \mu$s i.e. $\Omega_e/2\pi=90$ MHz we can achieve $99.35\%$ fidelity for $l=9.3\ \mu$m. 

\subsection{Blockade error}
\begin{figure}	
  \centering
 \includegraphics[width=4.2cm]{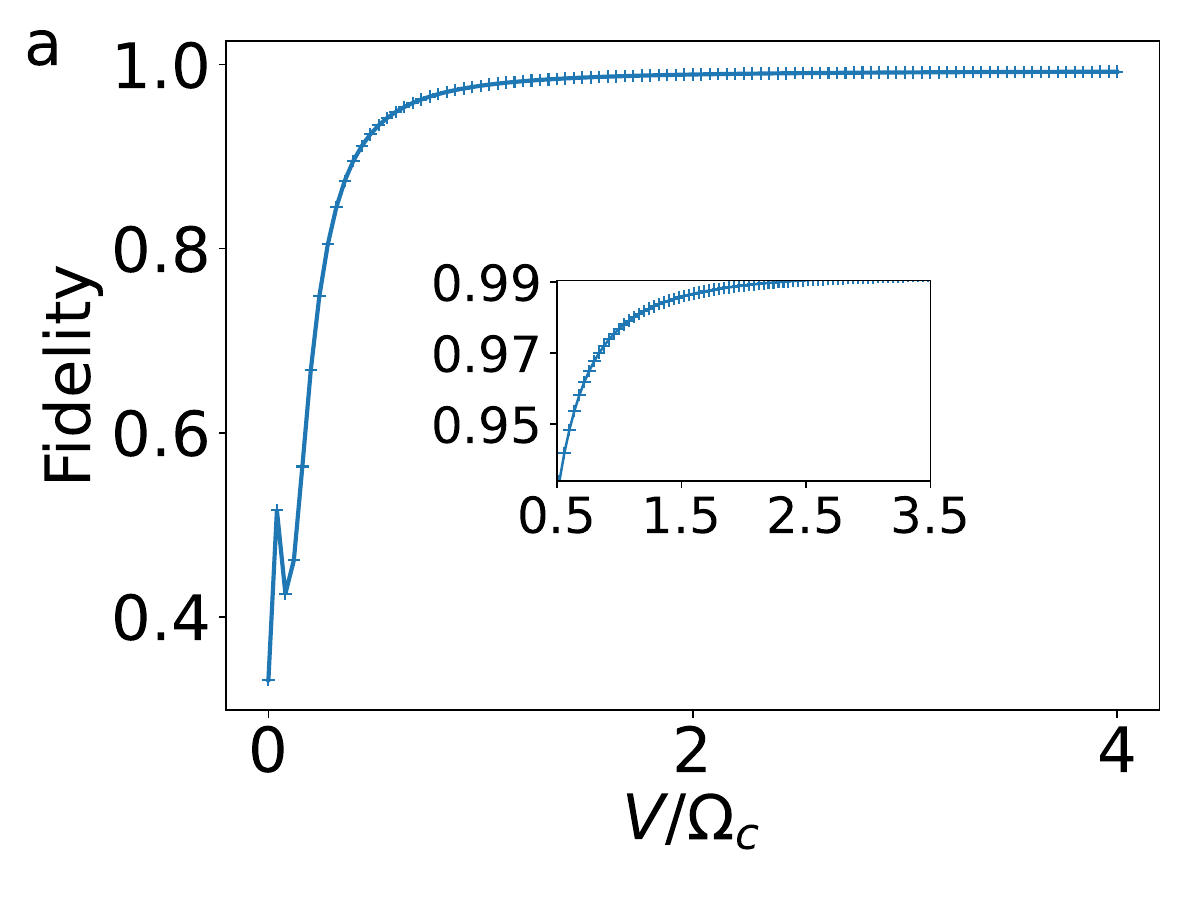}
 \includegraphics[width=4.2cm]{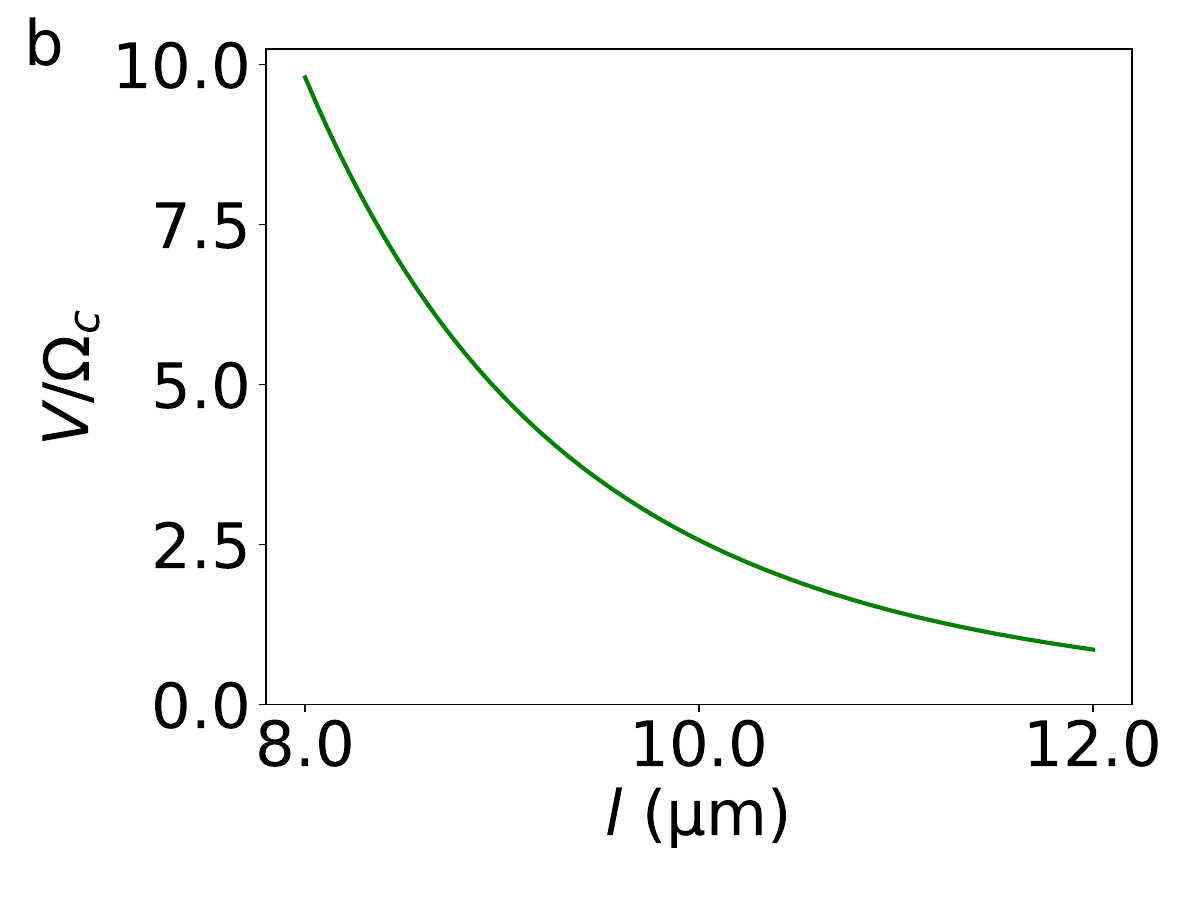}
 \\
 \caption{(a) For $V> 2.5\Omega_c$ more than 99\% fidelity can be achieved, while we fix $\Omega_c=2.5\Omega_e$. (b) The vdW interaction strength is plotted as a function of inter-atomic distance.}
 \label{fidv}
\end{figure}
In Rydberg atom quantum gates, blockade error arises when the Rydberg blockade mechanism is not perfect. Ideally, excitation of one atom to a Rydberg state should shift the energy levels of nearby atoms strongly enough to completely prevent their excitation. However, in practice, the interaction strength is finite, and the laser Rabi frequency may be comparable to or larger than the blockade shift. As a result, there is a small probability that two atoms are simultaneously excited to Rydberg states. The blockade error typically scales as $(\Omega_c/V)^2$. For conventional Rydberg blockade-based gates, the interaction strength must significantly exceed the Rabi frequency in order to suppress blockade errors. For our protocol we show in Fig.\ref{fidv}a we show that $>99\%$ fidelity can be achieved for $V>2.5\Omega_c$ that proves our proposed scheme is blockade error resilient.
\subsection{Applications}
\subsubsection{Deutsch–Jozsa algorithm}
\begin{figure} 
    \centering 
    \includegraphics[width=7.9cm]{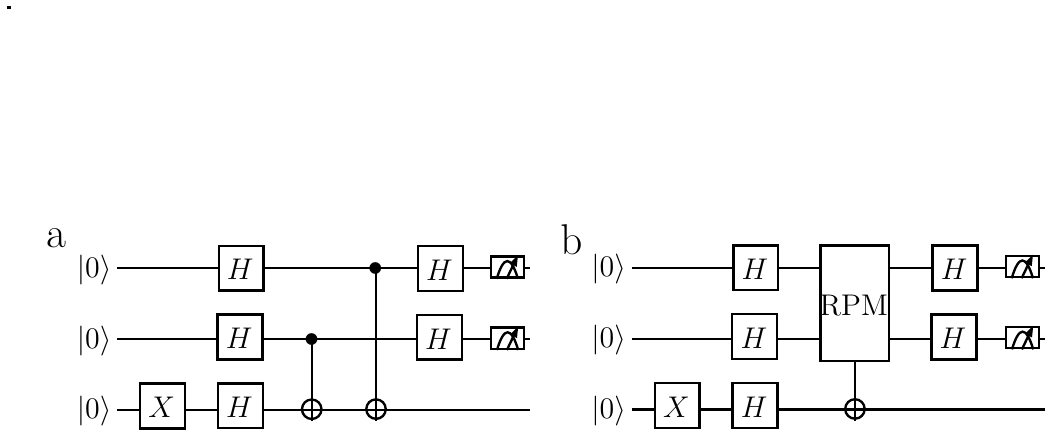}\\
     \includegraphics[width=8cm]{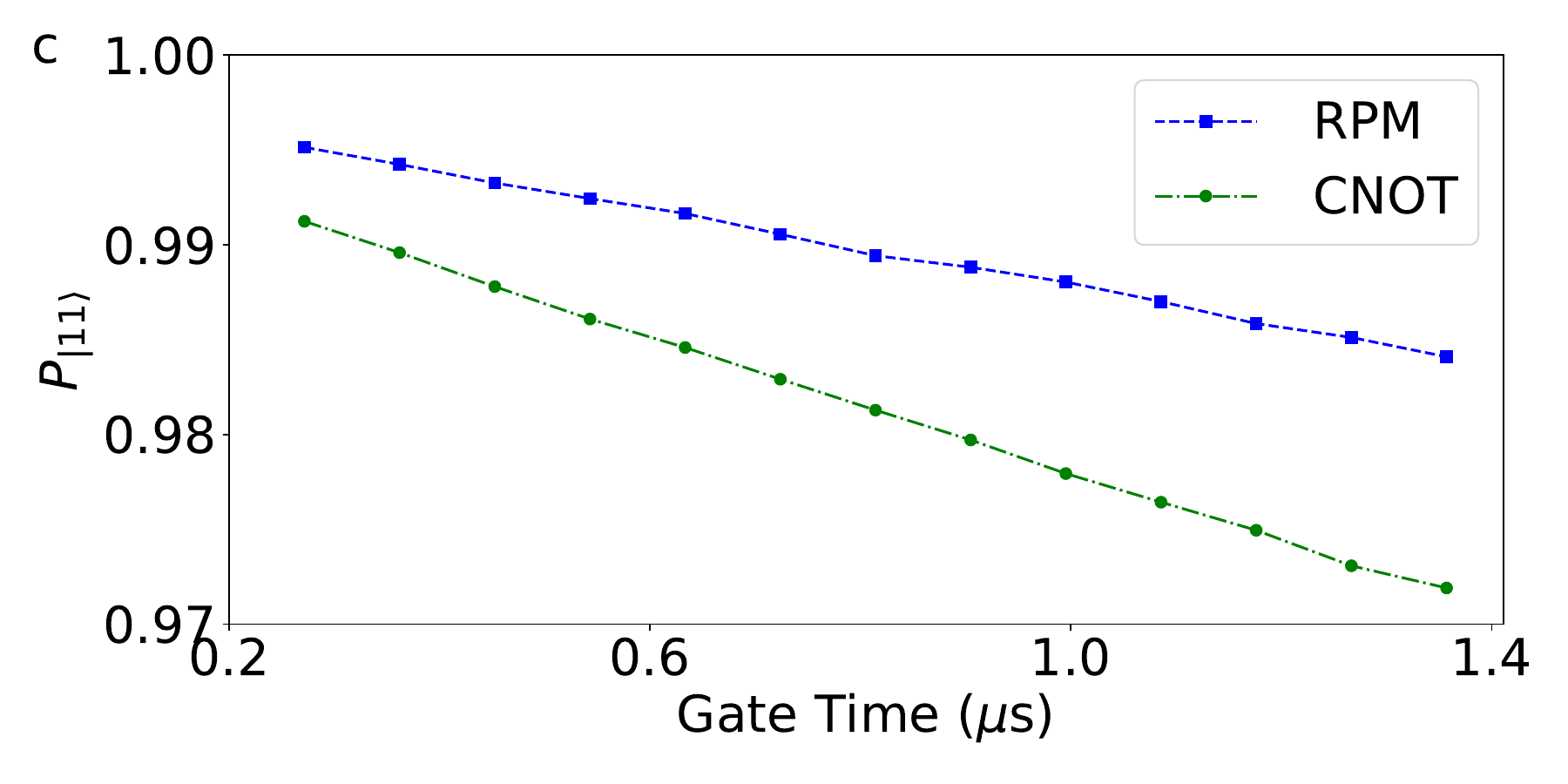}
    \caption{(a) an (b) are the circuits for simulating the Deutsch–Jozsa algorithm using CNOT gates and RPG, respectively, for the function  $f(x_i)=x_1\oplus x_2$. (c) shows how the probability of getting $\ket {11}$ (i.e., detecting as a balanced function) changes with changing individual gate operation time. }
    \label{fig4}
\end{figure}

Deutsch–Jozsa algorithm \cite{DeutschJozsa1992} is a very basic example of a quantum algorithm that shows the supremacy of quantum computation over its classical counterpart. The main target of this algorithm is to identify if a function with $n$ variables is balanced or constant. Classically, to determine this, one must do $(n/2+1)$ operations, but quantum mechanically, we just need one operation to perform. Thus, it reduces the total computation time as well as computational resources. For example, here we choose a function $f(x_i)=x_1\oplus x_2$ to examine if it is a balanced or constant function. We simulate the Deutsch–Jozsa algorithm using the conventional EIT-based CNOT gates and our proposed RPG using the circuit shown in the fig.\ref{fig4} a and b, respectively. Getting $\ket{11}$ as output indicates the function is balanced, or if we get $\ket{00}$ we confirm that the function is constant. Fig \ref{fig4} c shows how the probability of getting $\ket{11}$ in output varies with individual gate operation time.  Clearly, we get better results using RPG than using the CNOT gate, as using RPG obviously reduces circuit depth to reduce circuit noise.
\subsubsection{DQS}
\begin{figure} [t]
    \includegraphics[width=8.1cm]{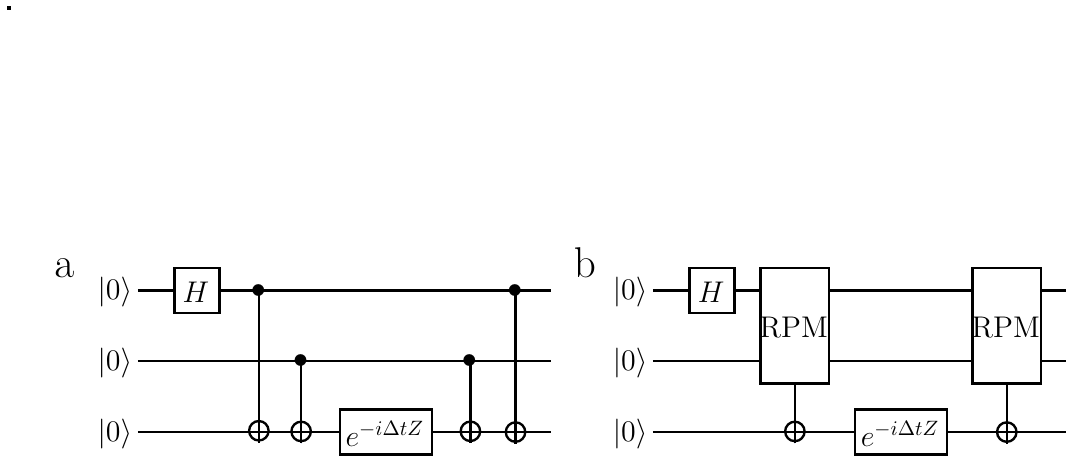}\\
     \includegraphics[width=8cm]{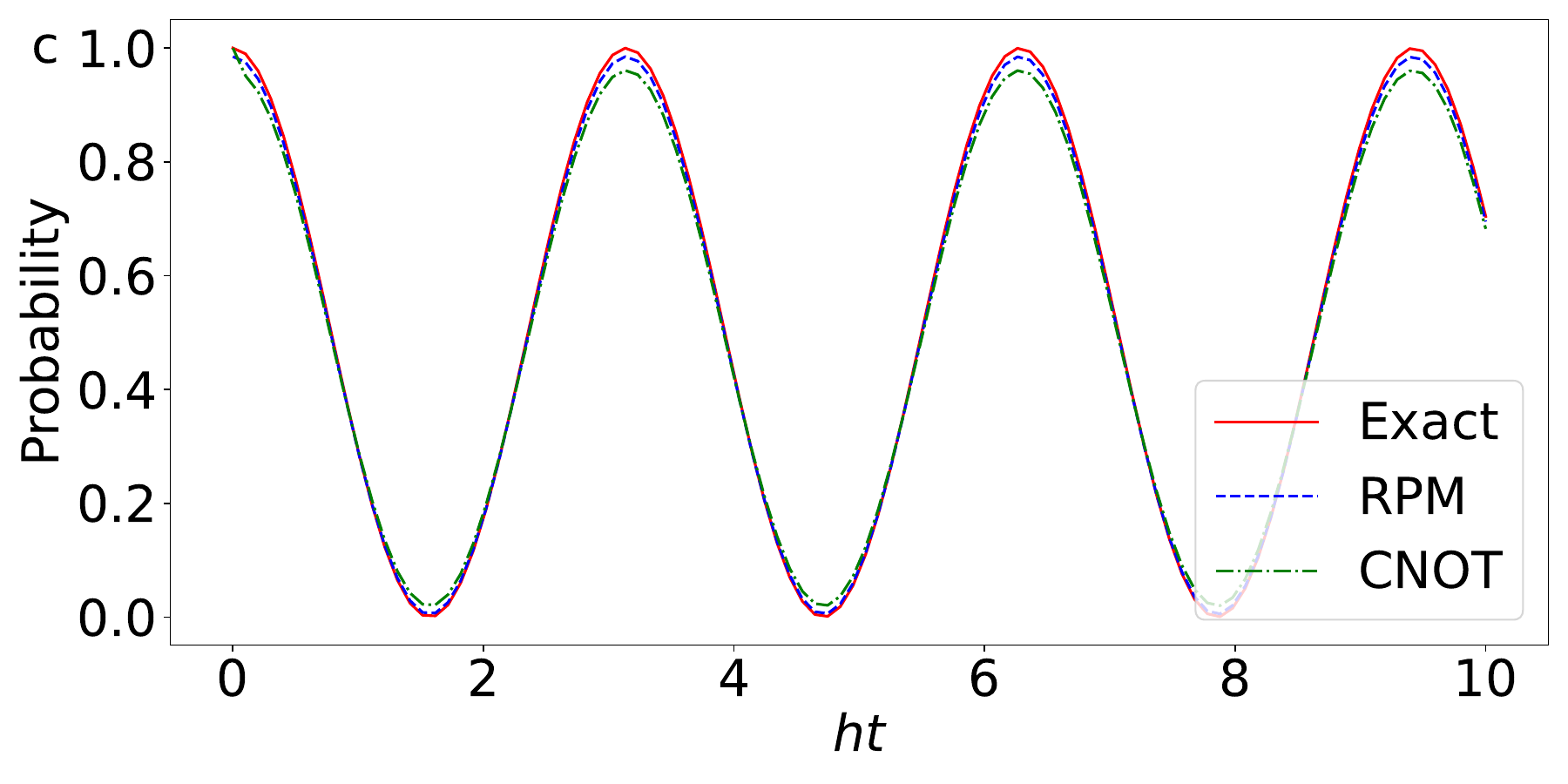}
    \caption{ (a) and (b) are the circuits to simulate the Hamiltonian $H=hZ_1Z_2$ for a two-qubit system, with the initial state $\ket{\psi_i}=1/\sqrt{2}(\ket{00}+\ket{10})$ using CNOT gates and RPG, respectively. (c) The time evolution of the state $\ket{\psi_i}$ is plotted; the RPG-based result resembles the exact one more closely than the CNOT-based result.}
    \label{fig5}
\end{figure}
 Digital quantum simulation uses a programmable quantum computer to simulate the dynamics of another quantum system by breaking the simulation into a sequence of quantum gates \cite{Berry2015, NielsenChuang}. For example, here we choose to simulate the very basic Ising Hamiltonian $H=hZ_1Z_2$ for a two-qubit system. We take a non-eigenstate $\ket{\psi_i}=1/\sqrt{2}(\ket{00}+\ket{10})$ as the initial state so that we can expect an evolution dynamics. Fig. \ref{fig5}a and b show the circuit to simulate the above-mentioned Hamiltonian starting with $\ket{\psi_i}$ using EIT-based CNOT gates and our proposed RPG, respectively. For instance, we choose the total gate operation time $0.8\ \mu$s for both the CNOT gate and RPG for simulation. Fig. \ref{fig5}c shows that the evolution plot (i.e. $|{\braket{\psi_i|\psi(t)}}|^2$ vs $ht$) using the RPG-based circuit is closer to the exact one than using the CNOT-based circuit.
\section{Acknowledgment}
We gratefully acknowledge Prof. Mark Saffman for insightful discussions and valuable suggestions.
\section{Conclusions}\label{cncl}
In conclusion, we have proposed an alternative scheme for QC and DQS using RPG instead of CNOT or CZ gates. We have constructed a three-qubit RPG based on dark state resonances and analyzed the time evolution of the computational basis by varying gate parameters. We have shown that the fidelity of our proposed RPG can be as high as $99.35 \%$ for the realistic experimental parameters. Moreover, our RPG is robust against blockade error. Using the same scheme, one can construct an n-qubit RPG for more than two control qubits by just addressing the shift in the target atom Rydberg state by applying additional lasers with some detuning as per requirement. Further, to show the utility of RPG as a substitute of the CNOT gate, we have simulated the Deutsch–Jozsa algorithm and two-qubit Ising Hamiltonian. It has been observed that employing the dark state-based RPG yields superior results compared to the conventional dark state-based CNOT gate, owing to the reduction of both circuit depth and circuit noise.

\bibliography{ref}
\end{document}